\documentclass[aps,a4paper,pra,floats,amsfonts,showkeys,showpacs,preprint,fleqn]{revtex4}

\usepackage{amsmath}
\usepackage{amsfonts}
\usepackage{amssymb}
\usepackage{bm}

\newcommand{\bd}{\begin{displaymath}}
\newcommand{\ed}{\end{displaymath}}
\newcommand{\be}{\begin{equation}}
\newcommand{\ee}{\end{equation}}
\newcommand{\bea}{\begin{eqnarray}}
\newcommand{\eea}{\end{eqnarray}}
\newcommand{\bean}{\begin{eqnarray*}}
\newcommand{\bra}[1]{\langle#1|}
\newcommand{\ket}[1]{|#1\rangle}
\newcommand{\eean}{\end{eqnarray*}}

\begin{document}

  \title{An Improved Tripartite Bell-type Inequality}
  \author{Fatemeh Shirdel}
  \email{fatemeh_shirdel@catholic.org}
  \author{Hossein Movahhedian}
  \email{hossein_movahhedian@catholic.org}
  \affiliation{Department of Physics, Shahrood University of Technology,
    Seventh Tir Square, Shahrood, Iran}

  \begin{abstract}
    So far, various Bell type inequalities have been introduced to test the
    local realism in tripartite systems. In this article we consider a
    tripartite
    system with two measurements in each side and two outputs for each
    measurement. Then we will introduce a new bell type inequality for this
    system and we show that this inequality is violated by quantum theory
    with a  violation factor and amount of violation of 3.5 and 2.5 respectively,
    which exceed those of available inequalities in both cases.
    Also we will show that the white noise tolerance of this
    new Bell type expression is 0.5, which agrees with the maximum amount of white
    noise tolerance of available inequalities up to now.
    \begin{center} \today \end{center}
  \end{abstract}


\maketitle

\section{Introduction \label{sec_1}}
  Using local theory, John S. Bell introduced an inequality which is violated
  by quantum theory. Later experiments showed that quantum theory is
  basically non-local \cite{BEL_J_S_64}.

  As the non-locality feature of quantum theory is intensively used in quantum
  information, Bell type inequalities have received more attention in recent
  years \cite{EKE_A_91}.

  Bell original inequality did not have any capability to be studied empirically in the
  laboratories. After that, Clauser, Horne, Shimony and Holt introduced their
  famous inequality called 
  CHSH which was reconsidered in laboratories since then. 
  As no experiment is error-free, there was
  an endeavor to gain a kind of Bell type inequality that would be violated
  as much as possible, so that it would be experimentally
  easy to test non-locality feature of quantum theory \cite{CHSH_69}.

  Svetlichny (in 1987) and Mermin (in 1990) obtained inequalities for
  tripartite systems which implied stronger violation of local
  theories \cite{SVE_G_87,MER_N_90}. Also in 1989,
  Greenberger, Horn, and Zeilinger obtained some inequalities for
  N-Particle systems ($N>2$) \cite{GHZ_89}.

  In this article, we introduce a new Bell type expression for 
  tripartite systems with two measurements in each side and two outputs for
  each measurement.
  Then, we will show that the violation factor (i.e. the ratio of the value
  of Bell expression according to quantum theory to its value according to
  local theory) and the amount of violation (i.e. the difference between the value
  of Bell expression according to quantum theory and its extermum value 
  according to local theory) of
  this inequality exceed those of available inequalities
  \cite{SVE_G_87,MER_N_90,BE_A_KL_D_93}, while its white noise tolerance agrees
  with the previous results.

\section{Tripartite systems \label{sec_2}}
  We consider a tripartite system consisting of particles $\mathcal{A}$,
  $\mathcal{B}$ and $\mathcal{C}$.

  Two possible measurements  $A$ and $A'$ are performed on particle $\mathcal{A}$ with
  outputs $a$ and $a'$ respectively, $B$
  and $B'$ on particle $\mathcal{B}$ with outputs $b$
  and $b'$ respectively, and finally $C$ and $C'$ on particle
  $\mathcal{C}$ with outcomes $c$ and $c'$ respectively, where $a,a',b,b',c,c' \in \{0,1\}$.

  Let $P_L(A,A',B,B',C,C'|a,a',b,b',c,c')$ denotes the triple joint probability
  that  measurements $A$, $A'$ on particle $\mathcal{A}$ result $a$ and $a'$
  respectively,
  measurements $B$, $B'$ on particle
  $\mathcal{B}$ result $b$ and $b'$ respectively, and measurements $C$ and
  $C'$ on particle $\mathcal{C}$ result
  $c$ and $c'$ respectively.

  It is obvious that:
  \be
    \sum_{a,a'}\sum_{b,b'}\sum_{c,c'} P_L(A,A',B,B',C,C'|a,a',b,b',c,c') = 1
    \label{eq_1}
  \ee
  Also let $P(A,B,C|a,b,c)$ denotes the joint probability that measurement A on particle
  $\mathcal{A}$ results "$a$", measurement $B$ on particle
  $\mathcal{B}$ results "$b$", and measurement $C$ on particle
  $\mathcal{C}$ results "$c$".

  Clearly:
  \be
    P(A,B,C|a,b,c)=
    \sum_{a'}\sum_{b'}\sum_{c'} P_L(A,A',B,B',C,C'|a,a',b,b',c,c')
    \label{eq_2}
  \ee			
  The normalization of $P$'s implies:
  \be
    \sum_{a}\sum_{b}\sum_{c} P(A,B,C|a,b,c)=1
    \label{eq_3}
  \ee
  As it is well known, a Bell type expression, $\Bbb B$, is a linear combination of
  joint probabilities that is bounded by local theories, i.e.
  \be
    \Bbb{B}=\sum_{I,J,K,l,m,n}\gamma(I,J,K | l,m,n) P(I,J,K | l,m,n)
    \label{eq_4}
  \ee
  where $I\in \{A,A'\},J\in \{B,B'\},K\in \{C,C'\},
            l\in \{a,a'\},m\in \{b,b'\}$ and $n\in \{c,c'\}$.
  Using equation (\ref{eq_2}) the Bell inequality in terms of $P_L$'s would become:
  \be
    \Bbb{B}=\sum_{a,a'}\sum_{b,b'}\sum_{c,c'} [\alpha(a,a',b,b',c,c')-
       \beta(a,a',b,b',c,c')] P_L(A,A',B,B',C,C'|a,a',b,b',c,c').
    \label{eq_5}
  \ee
  It is clear that
  \be
    -e \le \Bbb{B} \le f
    \label{eq_6}
  \ee
  where $f$ ($e$) is the greatest of positive real numbers $\alpha$'s ($\beta$'s) in equation (\ref{eq_5}).

\section{A New Bell Expression
         \label{sec_3}}
  One of the well-know Bell type expressions for tripartite systems is Mermin
  inequality, which can be expressed as \cite{CKH_08}:
  \be
    M=| E(A,B',C') + E(A',B,C') + E(A',B',C) - E(A,B,C) |
    \label{eq_7}
  \ee
  where
  \be
    E(A,B,C) = \langle A,B,C \rangle =
      \sum_{a}\sum_{b}\sum_{c} (-1)^z P(A,B,C|a,b,c)
    \label{eq_8}
  \ee
  and $P(A,B,C|a,b,c)$ is the joint probability discussed above. In the above
  equation $"z"$ is the number of zero's resulted in each particular
  setting \cite{SVE_G_87}.

  It is shown in \cite{MER_N_90} that Mermin inequality for local theories satisfies
  \be
    0 \le M \le 2
    \label{eq_9}
  \ee

  However, according to quantum theory, the upper bound of Mermin inequality is
  $4$ which shows that
  quantum theory is non-local. Here, the violation factor and 
  amount of violation in Mermin inequality are both 2 and the
  maximum white noise tolerance calculated is 0.5 \cite{CKH_08}.

  Now let's consider the following inequality for a tripartite system
  \bea
    G & = &  P(A,B,C|1,1,1)     + 5P(A,B,C|1,0,0)    + 5P(A,B,C|0,0,1)    +
             \nonumber \\
      &   &  P(A,B,C|1,0,1)     + 4P(A,B,C|0,0,0)    + 4P(A,B,C|0,1,0)    +
             \nonumber \\
      &   &  P(A,B',C'|0,0,0)   + P(A,B',C'|0,1,1)   - 4P(A,B',C'|0,0,1)  -
             \nonumber \\
      &   & 4P(A,B',C'|0,1,0)   -  P(A',B',C|0,0,1)  -  P(A',B',C|1,1,1)  -
             \nonumber \\
      &   &  4P(A',B',C|0,1,0)  - 4P(A',B',C|1,0,0)  - 5P(A',B,C'|1,0,0)  -
             \nonumber \\
      &   &  5P(A',B,C'|0,0,1)  +  P(A',B',C'|1,1,0) +  P(A',B',C'|0,0,1) -
             \nonumber \\
      &   &  4P(A',B',C'|1,1,1) - 4P(A',B',C'|0,0,0)
    \label{eq_10}
  \eea
  In appendix \ref{app_1}, it is shown that
  \be
    G \le 1.
    \label{eq_11}
  \ee
  However for a three-qubit Greenberger-Horne-Zeilinger state~\cite{GHZ_89} which is
  \be	
    \ket{\Psi}_{GHZ} =\frac{1}{\sqrt{2}}
      (\ket{\uparrow\uparrow\uparrow}_z +
       \ket{\downarrow\downarrow\downarrow}_z),
    \label{eq_12}
  \ee
  where $\uparrow$ and $\downarrow$ are spin polarization along z axis, and
  if $A=\sigma_X^A$, $A'=\sigma_Y^A$, $B=\sigma_X^B$,
  $B'=\sigma_Y^B$, $C=\sigma_X^C$ and $C'=\sigma_Y^C$, $G$ would become
  \be
    G=\frac{1}{4} + \frac{5}{4} + \frac{5}{4} + 0 + 0+ \frac{4}{4} + \frac{1}{4}
      + \frac{1}{4} - 0 - 0 - 0 - 0 - 0 - 0 - 0 - 0 + \frac{1}{8} + \frac{1}{8}
      - \frac{4}{8} - \frac{4}{8} = \frac{7}{2}
    \label{eq_13}
  \ee
  It is seen that the violation factor and the amount of violation of the inequality (\ref{eq_10}) 
  are 3.5 and 2.5 respectively, whereas the maximum
  violation factor and maximum amount of violation of the available inequalities so far, are 2.

  To calculate the white noise tolerance of G in tripartite
  systems, we consider the following density matrix:
  \be
    \rho = (1-p)\ket{\Psi}_{GHZ}~{_{GHZ}}\bra{\Psi} + \frac{p}{8}I.
    \label{eq_14}
  \ee
  Obviously
  \be
    P(A,B,C|a,b,c) = \frac{p}{8} + (1-p)
    P_{QM}(A,B,C|a,b,c)
    \label{eq_15}
  \ee
  where $P_{QM}(A,B,C|a,b,c)$ is the joint probability according to quantum theory and
  $p$ is the tolerance of Bell type expression. From equations
  (\ref{eq_10}) and (\ref{eq_15}), we have:
  \be
    p = \frac{G_{QM} - G_L}{G_{QM} - \frac{m-n}{8}}
    \label{eq_16}
  \ee
  where $G_{QM}$ is the value of our Bell expression, $G$, according to
  quantum theory, $G_L$ is its maximum value, according to local theories and 
  m(n) is the number of positive (negative) terms in $G$.

  It is easily seen that the white noise tolerance of G is 0.5, 
  which agrees with the maximum value calculated up to now.

\section{Conclusion \label{sec_4}}
  In this article we introduced a Bell type inequality for tripartite
  systems with two measurements for each side and
  two outputs for each measurement, which is 
  violated by quantum theory with a stronger violation factor and more amount
  of violation
  than the available inequalities. In fact, the violation factor and the amount of
  violation
  of our inequality
  are 3.5 and 2.5 respevtively,
  which are 1.5 and 0.5 more than the results obtained so far, respectively.
  However the tolerance of our inequality is the
  same as others. This increment of violation factor and the amount of
  violation
  increase the accuracy of
  experiments in which the errors are inevitable. 
  
  Also one of the advantages of our inequality is that it includes only 20 different
  joint probabilities whereas in other works it is much more than this (in
  Mermin and Svetlichny inequalities it is 32 and 64 respectively).
  So, our inequality requires less measurements which in turn, reduces the errors
  due to experiment. See \cite{MOV_H_09}. 

\appendix

\section{List of Joint Probabilities \label{app_1}}
  In this appendix we derive equation (\ref{eq_11}). From the definition
  (\ref{eq_2}) and denoting $P_L(A,A',B,B',C,C'|a,a',b,b',c,c')=P_{aa'bb'cc'}$
  for simplicity, we have
  \bean
    P(A,B,C|1,1,1) & = & P_{101010} + P_{101011} + P_{101110} + P_{101111} + \\
                     & & P_{111010} + P_{111011} + P_{111110} + P_{111111}   \\
    P(A,B,C|1,0,0) & = & P_{100000} + P_{100001} + P_{100100} + P_{100101} + \\
                     & & P_{110000} + P_{110001} + P_{110100} + P_{110101}   \\
    P(A,B,C|0,0,1) & = & P_{000010} + P_{000011} + P_{000110} + P_{000111} + \\
                     & & P_{010010} + P_{010011} + P_{010110} + P_{010111}   \\
    P(A,B,C|1,0,1) & = & P_{100010} + P_{100011} + P_{100110} + P_{100111} + \\
                     & & P_{110010} + P_{110011} + P_{110110} + P_{110111}   \\
    P(A,B,C|0,0,0) & = & P_{000000} + P_{000001} + P_{000100} + P_{000101} + \\
                     & & P_{010000} + P_{010001} + P_{010100} + P_{010101}   \\
    P(A,B,C|0,1,0) & = & P_{001000} + P_{001001} + P_{001100} + P_{001101} + \\
                     & & P_{011000} + P_{011001} + P_{011100} + P_{011101}   \\
    P(A,B',C'|0,0,0) & = & P_{000000} + P_{000010} + P_{001000} + P_{001010} + \\
                     & & P_{010000} + P_{010010} + P_{011000} + P_{011010}   \\
    P(A,B',C'|0,1,1) & = & P_{000101} + P_{000111} + P_{001101} + P_{001111} + \\
                     & & P_{010101} + P_{010111} + P_{011101} + P_{011111}   \\
    P(A,B',C'|0,0,1) & = & P_{000001} + P_{000011} + P_{001100} + P_{001011} + \\
                     & & P_{010001} + P_{010011} + P_{011001} + P_{011011}   \\
    P(A,B',C'|0,1,0) & = & P_{000100} + P_{000110} + P_{001100} + P_{001110} + \\
                     & & P_{010100} + P_{010110} + P_{011100} + P_{011110}   \\
    P(A',B',C|0,0,1) & = & P_{000010} + P_{000011} + P_{001010} + P_{001011} + \\
                     & & P_{100010} + P_{100011} + P_{101010} + P_{101011}   \\
    P(A',B',C|1,1,1) & = & P_{010110} + P_{010111} + P_{011110} + P_{011111} + \\
                     & & P_{110110} + P_{110111} + P_{111110} + P_{111111}   \\
  \eean
  \bean
    P(A',B',C|0,1,0) & = & P_{000100} + P_{000101} + P_{001100} + P_{001101} + \\
                     & & P_{100100} + P_{100101} + P_{101100} + P_{101101}   \\
    P(A',B',C|1,0,0) & = & P_{010000} + P_{010001} + P_{011000} + P_{011001} + \\
                     & & P_{110000} + P_{110001} + P_{111000} + P_{111001}   \\
    P(A',B,C'|1,0,0) & = & P_{010000} + P_{010010} + P_{010100} + P_{010110} + \\
                     & & P_{110000} + P_{110010} + P_{110100} + P_{110110}   \\
    P(A',B,C'|0,0,1) & = & P_{000001} + P_{000011} + P_{000101} + P_{000111} + \\
                     & & P_{100001} + P_{100011} + P_{100101} + P_{100111}   \\
    P(A',B',C'|1,1,0) & = & P_{010100} + P_{010110} + P_{011100} + P_{011110} + \\
                     & & P_{110100} + P_{110110} + P_{111100} + P_{111110}   \\
    P(A',B',C'|0,0,1) & = & P_{000001} + P_{000011} + P_{101001} + P_{101011} + \\
                     & & P_{100001} + P_{100011} + P_{111110} + P_{111111}   \\
    P(A',B',C'|0,0,0) & = & P_{000000} + P_{000010} + P_{001000} + P_{001010} + \\
                     & & P_{100000} + P_{100010} + P_{101000} + P_{101010}   \\
    P(A',B',C'|1,1,1) & = & P_{010101} + P_{010111} + P_{011101} + P_{011111} + \\
                     & & P_{110101} + P_{110111} + P_{111101} + P_{111111}.
  \eean
  Inserting the above joint probabilities in equation (\ref{eq_10}), we get
  \bean
    G & = &  P_{000000} +  P_{100000}  +  P_{001000} - 4P_{010000} -  4P_{000100} + \\
      &   &  P_{000010} -  4P_{000001} -  4P_{110000} -  4P_{101000} +  P_{100100} - \\
      &   &  4P_{100010} +  P_{100001} +  P_{011000} -  4P_{010100} +  P_{010010} - \\
      &   &  4P_{010001} -  4P_{001100} -  4P_{001010} +  P_{001001} +  P_{000110} - \\
      &   &  4P_{000101} -  4P_{000011} -  4P_{111000} +  P_{110100} -  4P_{110010} + \\
      &   &  P_{110001} +  P_{011100} +  P_{011010} -  4P_{011001} -  4P_{001110} + \\
      &   &  P_{001101} +  P_{000111} -  4P_{111111} -  4P_{011111} +  P_{101111} - \\
      &   &  4P_{110111} +  P_{111011} - 4  P_{111101} +  P_{111110} +  P_{001111} + \\
      &   &  P_{010111} -  4P_{011011} +  P_{011101} -  4P_{011110} -  4P_{100111} + \\
      &   &  P_{101011} -  4P_{101101} +  P_{101110} +  P_{110011} +  P_{110101} - \\
  \eean
  \bean
      &   &  4P_{110110} -  4P_{111001} +  P_{111010} +  P_{111100} -  4P_{101100} - \\
      &   &  4P_{001011} -  4P_{100011} -  4P_{100101} +  P_{100110} -  4P_{101010} + \\
      &   &  P_{010101} +  P_{010011} -  4P_{010110} +  P_{101001},
  \eean
  which according to equation \ref{eq_11} is less than or equal to 1. Please note that
  all P's are positive here.

\end{document}